\documentclass[a4paper]{article}
\usepackage{amsmath}
\usepackage{amssymb}
\begin{document}
\title{\bf Self-consistent model of fermions}
\author{Vladimir Yershov \\
{\small \it Mullard Space Science Laboratory} \\ 
{\small \it (University College London),} \\ 
{ \small \it  Holmbury St.Mary, Dorking RH5 6NT, UK} \\
{ \small vny@mssl.ucl.ac.uk}}
\date{}
\sloppy
\maketitle
\begin{abstract}
A composite model of the fundamental fermions based on colour preons
is discussed. It is found that, if endowed with the pairwise 
(repulsive/attractive) chromoelectric fields,
preons would cohere in a series of structures, resembling
by their properties the three generations of the fundamental fermions. 
The model is self-consistent in the sense that it makes no use of free 
or experimental input parameters.
\end{abstract}

\newcommand{\abs}[1]{\lvert#1\rvert}

\section{Introduction}
The Standard Model of particle physics does
not explain the hierarchical pattern observed in the masses
of quarks and leptons because it uses these masses as
its input parameters and it is not aimed at answering 
the question as to why the masses are distributed in 
the particular way they are. At a first glance they seem to be random, 
except for the fact that the masses generally increase with 
the generation number.

Numerous attempts to solve this puzzle have been made over
the last twenty years, but the problem is still there. The
observed pattern of the particle masses insistently points 
to structures beyond the quark scale. A number of models
for these hypothetical structures have been proposed, focusing 
on the ideas of unification \cite{georgi}, technicolor 
\cite{weinberg}, supersymmetric unification \cite{dimopoulos},
strings and branes \cite{green}. The latter approach gives
some encouraging results, such as distributions resembling
the observed families of particles, but no more. 

There exist also models where the 
fundamental particles are composite entities \cite{luty}, but 
these models are not very popular, being far from reality and
facing problems with resolving the mass paradox and meeting  
the 't Hooft anomaly matching constraints. 
Finally, there are models discussing the possibility of 
randomness of these masses \cite{donoghue}, which could easily
be falsified by constructing a counter-example where 
the fermion masses would be functionally dependent. 
For example, G.R.Filewood proposed a model \cite{filewood}, 
regarding particles as almost crystalline symmetric
structures, predicting some fermion masses with 
based on the geometrical approach to the structures 
a fairly good accuracy. This shows that at least some of the  particle 
masses have a common origin. Here we shall discuss yet another 
composite model based on sub-quark primitive particles usually called 
preons. In author's view, this model is closer to reality -- at least it 
correctly reproduces the masses of all three generations of the 
fundamental fermions.    

\section{Simple structures}

Let us suppose that the preon  $\Pi$ (the basic building block of the composite 
fermions) could be regarded as a source of a spherically-symmetric pairwise field 
$F$ with the colour (tripolar) symmetry, $\Pi, \overline{\Pi} \in \bf{3}_c$. 
Except for its tripolarity, the field $F$ is analogous to the Lennard-Jones 
fields used in molecular physics for modelling long-range attractive 
($\phi_s$) and short-range repulsive ($\phi_e$) forces. A trade-off 
between these forces leads to equilibrium configurations of basic particles.
Assuming that infinite energies are not accessible in nature we can
hypothesise that the energy of both $\phi_s$ and $\phi_e$, after 
reaching the maximum, decays to zero at the origin. The simplest form 
for such a pairwise field would be:
\begin{equation}
\begin{split}
& F=\phi_s+\phi_e, \\
& \phi_s=s\exp{(-\rho^{-1})}, \hspace{0.5cm} \phi_e=-\phi'_s(\rho),
\end{split}
\label{eq:basicfield}
\end{equation}
\noindent
where the signature $s=\pm 1$ indicates the sense of the interaction (attraction
or repulsion); the derivative of $\phi_s$ is taken with respect to the radial
coordinate $\rho$. Far from the source, the second component of the field $F$ mimics 
the Coulomb gauge, whereas the first component extends to infinity being
almost constant (similarly to the strong field).
Let us formally represent the preon by using a set of auxiliary 
$3\times 3$ singular matrices ~$\Pi^i$ with the following elements:
\begin{equation}
^{\pm}\pi^i_{jk}=\pm\delta^i_j (-1)^{\delta^k_j},
\label{eq:pmatrix}
\end{equation}
where $\delta^i_j$ is the Kronecker delta-function;
the $\pm$-signs correspond to the sign of the charge; 
and the index $i$ stands for the colour ($i=1,2,3$ or red, green and blue).
The diverging components of the field can be represented by 
reciprocal elements:
$\tilde{\pi}_{jk}=\pi_{jk}^{-1}$.
Then, we can define the preon's (unit) charges and masses 
by summation of these matrix elements:  
\begin{equation}
\begin{split}
q_{\Pi}&=\mathbf{u}^{\intercal}{\Pi}\mathbf{u} \hspace{0.1cm},
 \hspace{0.3cm} \tilde{q}_{\Pi}=\mathbf{u}^{\intercal} \tilde{\Pi}
\mathbf{u} \\
m_{\Pi}&=\abs{\mathbf{u}^{\intercal}{\Pi}\mathbf{u}} \hspace{0.1cm},
\hspace{0.3cm} \tilde{m}_{\Pi}=\abs{\mathbf{u}^{\intercal} \tilde{\Pi}
\mathbf{u}}
\end{split}
\label{eq:preonchargemass}
\end{equation}
($\mathbf{u}$ is the diagonal of a unit matrix; $\tilde{q}_\Pi$ 
and $\tilde{m}_\Pi$ diverge).   
By assuming that the field $F(\rho)$ does not act 
instantaneously at a distance we can define the mass 
of a system containing, say, $N$ preons, as 
proportional to the number of these particles, 
wherever their field flow rates are not cancelled.
For this purpose, we shall regard  the total 
field flow rate, $v_N$, of such a system as a 
superposition of the individual volume flow rates 
of its $N$ components. Then, the net mass of the system can 
be calculated (to a first order of approximation) as the number
of particles, $N$, times the normalised to unity (Lorentz-additive)
field flow rate $v_N$:
\begin{equation}
m_N=\abs{N v_N}.
\label{eq:mass}
\end{equation}
\noindent
Here $v_N$ is computed recursively from:
\begin{equation}
v_i=\frac{q_i+v_{i-1}}{1+\abs{q_i v_{i-1}}}, 
\label{eq:flowrate}
\end{equation}
with $i=2, \dots\,,N$ and putting $v_1=q_1$.
The normalisation condition (\ref{eq:flowrate}) 
expresses the common fact that the superposition
flow rate of, say, 
two antiparallel flows ($\uparrow \downarrow$) with equal
rate magnitudes 
$\abs{\mathbf{v}_\uparrow}=\abs{\mathbf{v}_\downarrow}=v$ vanishes 
($v_{\uparrow \downarrow}=0$), whereas, in the case of parallel
flows ($\uparrow \uparrow$) it cannot exceed the magnitudes of   
the individual flow rates ($v_{\uparrow \uparrow} \leq v$).
%
Then, when two unlike-charged preons combine (say, red and antigreen),
the magnitudes of their oppositely directed flow rates cancel each other
(resulting in a neutral system). The corresponding acceleration 
also vanishes, which is implicit in (\ref{eq:mass}).
This formula implies the complete cancellation of masses in the
system with vanishing electric fields (converted into the binding
energy of the system), but this is only an approximation 
because in our case the preons are separated from each other by 
distance of equilibrium, whereas the complete cancellation of flows 
is possible only when the flow source centres coincide.
%
Making use of the known pattern of attraction and repulsion
between colour charges 
(two like-charged but unlike-coloured particles are
attracted, otherwise they repel) we can write the signature $s_{ik}$
of the chromoelectric interaction between two preons with 
colours $i$ and $k$ as
\begin{equation}
s_{ik}=-\mathbf{u}^\intercal{\Pi}^i {\Pi}^k\mathbf{u}.
\label{eq:seforce}
\end{equation}
%
%
\noindent
This pattern implies that the preons of
different colours and charge polarities will cohere in structures, 
the simplest of which will be the charged
and neutral colour-doublets (dipoles)
\begin{equation*}
\varrho^\pm_{ik}= \mathbf{\Pi}^i+\mathbf{\Pi}^k, 
\end{equation*}
\begin{equation*}
\hspace{2.5cm}g^0_{ik}= \mathbf{\Pi}^i+\overline{\mathbf{\Pi}}^k, \hspace{0.6cm} i,k=1,2,3.
\end{equation*}

\noindent
According to (\ref{eq:preonchargemass}),
\begin{equation*}
q(\varrho_{ik})=\pm 2, \hspace{0.4cm}
m(\varrho_{ik})=2, \hspace{0.4cm}
\tilde{m}(\varrho_{ik})=\infty,
\end{equation*}
and 
\begin{equation*}
q(g_{ik})=0, \hspace{0.4cm}
m(g_{ik})=0, \hspace{0.4cm}
\tilde{m}(g_{ik})=\infty.
\end{equation*}
If an additional charged preon is added to the neutral
doublet, the mass and the charge of the system are restored: 
\begin{equation}
q(g_{ik})=\pm 1, 
\hspace{0.3cm}
m(g_{ik})=1,
\hspace{0.3cm}
\label{eq:mgluon}
\end{equation}
but still 
\begin{equation}
\tilde{m}(g_{ik})=\infty.
\label{eq:mprimgluon}
\end{equation}

The charged dipoles $\varrho$ ($2\Pi$ and  $2\overline{\Pi}$) 
cannot be free because of their diverging strong fields. 
Any distant preon of the same charge
but with a complementary colour will be attracted to the dipole,
thus, forming a (charged) tripole
denoted here as $y$ (and $\overline{y}$ for the opposite polarity): 
%
\begin{equation*}
y=\sum_{i=1}^3{\mathbf{\Pi}^i} \hspace{0.4cm} {\rm or} \hspace{0.4cm}
\overline{y}=\sum_{i=1}^3{\overline{\mathbf{\Pi}}^i}.
\end{equation*}
%
The tripole ($y$-particle) is colourless at infinity but 
colour-polarised nearby, which means that tripoles can
combine strings (pole-to-pole to each other) due to 
their residual chromaticism.
%
%
In order to formalise representation of these structures let
us introduce the following matrices:
\begin{equation*}
\alpha_0=
\begin{pmatrix} 1 & 0 & 0  \\ 
                0 & 1 & 0  \\
                0 & 0 & 1   
 \end{pmatrix}
 \text{,\hspace{0.5cm}}
\overline{\alpha}_0=-\frac{1}{2}   
\begin{pmatrix} 0 & 1 & 1  \\
                1 & 0 & 1  \\
                1 & 1 & 0 
\end{pmatrix},
\end{equation*}
\begin{equation*}
\alpha_1=
\begin{pmatrix} 0 & 1 & 0  \\ 
                0 & 0 & 1  \\
                1 & 0 & 0   
 \end{pmatrix}
 \text{,\hspace{0.5cm}}
\overline{\alpha}_1=-\frac{1}{2}   
\begin{pmatrix} 1 & 0 & 1  \\
                1 & 1 & 0  \\
                0 & 1 & 1 
\end{pmatrix}, 
\end{equation*}
\begin{equation*}
\alpha_2=
\begin{pmatrix} 0 & 0 & 1  \\ 
                1 & 0 & 0  \\
                0 & 1 & 0   
 \end{pmatrix}
 \text{,\hspace{0.5cm}}
\overline{\alpha}_2=-\frac{1}{2}   
\begin{pmatrix} 1 & 1 & 0  \\
                0 & 1 & 1  \\
                1 & 0 & 1 
\end{pmatrix}. 
\end{equation*}
Pairs of like-charged tripoles $y$ would combine in short
strings with their components $180^\circ$-rotated with respect to each other. 
The corresponding charged structure $\delta^\pm$ can be written as
\begin{equation*}
\delta^\pm = \alpha_i y + \overline{\alpha}_i y. 
\end{equation*}
with $q_\delta=\pm 6$,  $m_\delta=\tilde{m}_\delta=6$.
The states corresponding to
different $i$ are equivalent. 
The pairs of unlike-charged $y$-particles would also combine
(rotated by $\pm 120^\circ$ with respect to each other):
\begin{equation}
\gamma^0 = \alpha_i y + \alpha_k \overline{y} 
\label{eq:gamma}
\end{equation}
with $q_\gamma=0$, $m_\gamma=\tilde{m}_\gamma=0$, $i \neq k$,
$\gamma^0 \equiv 3(g\overline{g})$.

The $Z_3$-symmetry of the tripole $y$ implies that a string of three
like-charged tripoles (triplet) would close in a loop 
 $3y$ (or $3\overline{y}$) denoted here as $e$ because, 
as we shall see, this structure by its properties can be identified 
with the electron. The triplet $e$ is charged, with its charge $q_e=\pm 9$ and 
mass $m_e=9$ (expressed in units of preon's charge and mass).
The tripoles in this structure can be directed with 
their vertices towards or away from the centre 
of the loop.
However, these configurations correspond to two different 
phases of the same structure, since the tripoles here have a rotational 
degree of freedom (around their common ring-closed axis). 
At the same time, the tripoles
will orbit the centre of the structure moving along 
the ring-closed axis. The resulting currents have helical 
shapes with two possible helicity signs (clockwise or anticlockwise). 
These different helicities can be identified with two 
spins of the structure ($e_\uparrow$ and $e_\downarrow$).
%
The pairs of unlike-charged tripoles can form longer strings: 
\begin{equation*}
\nu_{e\uparrow}= 
\begin{pmatrix}
 \alpha_0 & \alpha_1 & \overline{\alpha}_1 & \overline{\alpha}_2  \\
 \alpha_2 & \alpha_0 & \overline{\alpha}_0 & \overline{\alpha}_1  \\
 \alpha_1 & \alpha_2 & \overline{\alpha}_2 & \overline{\alpha}_0  
\end{pmatrix}
\begin{pmatrix}
 y \\
 \overline{y} \\
 \overline{y} \\
 y
\end{pmatrix}   
\end{equation*}
or
\begin{equation*}
\nu_{e\downarrow}= 
\begin{pmatrix}
 \alpha_0 & \alpha_2 & \overline{\alpha}_2 & \overline{\alpha}_1  \\
 \alpha_1 & \alpha_0 & \overline{\alpha}_0 & \overline{\alpha}_2  \\
 \alpha_2 & \alpha_1 & \overline{\alpha}_1 & \overline{\alpha}_0  
\end{pmatrix}
\begin{pmatrix}
 y \\
 \overline{y} \\
 \overline{y} \\
 y
\end{pmatrix}.   
\end{equation*}
with the pattern of colour charges repeating after each six 
consecutive $y\overline{y}$-pairs, which allows the closure of
such a string in a loop (denoted here as 
$\nu_e=6y\overline{y}$). 
The structure $\nu_e$ (formed of 36 preons) is electrically neutral and 
has a vanishing mass, according to (\ref{eq:mass}), unless combined with 
a charged particle, say $y$ or $3y$, which would restore the entire
mass of the system.
%

\section{Combining $y$, $e$, and $\nu_e$}

Since the strong fields of $3y$ ($e$) and $6y\overline{y}$ 
($\nu_e$) are closed, these particles
can be found in free states. 
%
The particles $y$, $e$, and $\nu_e$ can combine with each other
because of their residual chromaticism.
The structure $y_1=y+\nu_e$  will have a mass
of 39 preon units ($m_{y_1}=n_\nu+m_y=36+3$)
and be charged, with its charge $q_{y_1}=\pm 3$, corresponding to the
charge of the $y$-particle. The charge of the configuration
$e+\nu_e$ will correspond to the charge of the triplet  
 \ $e$ ($q_{{\nu_e}e}=\pm 9$). Its mass will be 
 45 units as expressed in preon's units of mass ($n_\nu+m_e=36+9$).

%
\vspace{0.1cm}
Like-charged particles $y_1$ of the same helicity signs 
would further combine (through an intermediate 
$\nu_e$-particle with the opposite helicity) forming
three-component strings. 
The string $y_1\,e$\,\,$y_1$ can be identified with 
the {\it u}-quark. Its charge will correspond to the charge of 
two $y$-particles ($q_u=\pm 6$) and its mass will roughly
be the sum of the masses of its two $y_1$-components:
$m_u=2\times 39=78$ (preon mass units). The positively charged $u$-quark
(78 preon mass units) would be able to combine with the negatively charged particle 
$e^-\nu_e$ (45 preon mass units) mass, forming the {\it d}-quark with a mass
of 123 preon mass units, $m_d=m_u+m_{e\nu_e}=78+45=123$, 
 and with the charge derived from the charges of its constituents 
$q_d=q_u+q_e=+6-9=-3$ units.


\vspace{0.1cm}
The interactions between the particles $e$ and $\nu_e$ is 
be helicity-dependent. 
The configuration of colour charges of the structure 
$\nu_{e\uparrow}$ does not match that of the structure $e_\downarrow$,
which would lead to the mutual repulsion of these particles.
Only the structures $e_\uparrow$ + $\nu_{e\uparrow}$ 
or $e_\downarrow$ + $\nu_{e\downarrow}$ can be formed
since the combined potential of these structures implies attraction
between their components.
By contrast, if two structures of the same kind combine (e.g., $e$ with $e$ 
or $\nu_e$ with $\nu_e$), their helicity signs must 
be opposite in order to create an attractive force between the components of the pair.  
This coheres with and probably explains the Pauli exclusion principle, which
suggests identifying the helicities of the structures in question with their 
spins. 

\vspace{0.1cm}
Normalised to the number of its constituents (divided by nine) 
the charge of the $3y$-particle,
gives us the conventional unit charge of the electron.
Then, the charges of the $y$ and $\delta$-particles 
(fractions of the nine-unit electron charge)
would correspond to the quark fractional charges.

\vspace{0.1cm}
It is not unnatural to suppose that the particles of the second and third
generations of the fundamental fermions are formed of simpler structures 
belonging to the first generation.
For example, the muon-neutrino can be a bound state 
of the positively and negatively charged particles 
{$y_1$} and $\overline{y}_1$:
\begin{equation}
\nu_\mu=y\nu_{e\uparrow} \nu_{e\downarrow} \nu_{e\uparrow}\overline{y}
=y_1 \nu_e \overline{y}_1 , 
 \label{eq:numu}
\end{equation}
whereas the muon's structure can be written as
\begin{equation}
\mu= (\nu_{e\downarrow} e^-_\downarrow \overline{y})
(\nu_{e\uparrow} \nu_{e\downarrow}  \nu_{e\uparrow}y)
=\overline{\nu}_e e^-\nu_\mu .
\label{eq:muon}
\end{equation}
Similarly, the structures of the other higher-generation 
fermions can be found. It is difficult to regard these
structures as rigid bodies: they are rather oscillating clusters
with multiple resonances, whose oscillatory energies are likely 
to contribute to the masses of these systems.
In (\ref{eq:muon}) we have enclosed the supposedly clustered
components in parenthesis. 
%
Obtaining the masses of these systems is not 
a straightforward task, but one can show that, in principle, 
they are computable by using a simple empirical formula, which relates 
the oscillatory energies of the components to the sum $M$ of their 
masses $m_k$:

\begin{equation}
M=\sum_k{m_k} , 
\label{eq:summass}
\end{equation}

\noindent
multiplied by the sum $\tilde{M}$ of their reciprocal masses $1/\tilde{m}_k$:

\begin{equation*}
\frac{1}{\tilde{M}}=\sum_k{\frac{1}{\tilde{m}_k}} .
\end{equation*}
 
\noindent
Then, by setting for simplicity the unit-conversion coefficient to unity,
we can compute the masses of the combined structures as

\begin{equation*}
M_{\rm total}=M \tilde{M} 
\end{equation*}

\noindent 
We shall abbreviate this summation rule by using the overlined notation: 

\begin{equation}
M_{\rm total}=\overline{m_1+ m_2+ \ldots + m_N}=\frac{m_1+m_2+\dots+m_N}
{\tilde{m}_1^{-1}+\tilde{m}_2^{-1}+\dots+\tilde{m}_N^{-1}} .
\label{eq:mtotal}
\end{equation}

As an  example, let us compute the muon's mass. The masses of its 
components, according
to its structure, are: $m_1=\tilde{m}_1=48$, $m_2=\tilde{m}_2=39$ 
(in preon mass units). Then, the mass of the muon will be

\begin{equation*}
m_\mu =\overline{m_1+m_2} = \frac{m_1+m_2}{{\tilde{m}_1}^{-1}+
{\tilde{m}_2}^{-1}}=1872
 {\hspace{0.2cm} {\rm (preon \hspace{0.2cm} mass \hspace{0.2cm} units)}.}
\end{equation*}

\noindent
For the $\tau$-lepton, according to its structure,
 $m_1=\tilde{m}_1=156$, $m_2=\tilde{m}_2=201$, so that   

\begin{equation*}
m_\tau  = \overline{m_1+m_2}=31356 {\hspace{0.2cm} {\rm (preon 
\hspace{0.2cm}  mass \hspace{0.2cm} units)}.}
\end{equation*}

\vspace{0.1cm}
\noindent
For the proton, positively charged particle consisting of two $u$,
and one $d$ quarks submerged in a cloud of gluons $g_{ij}$, 
the masses of its components 
are $m_u=\tilde{m}_u=78$,  $m_d=\tilde{m}_d=123$,
$m_g=2m_u+m_d=279$, $\tilde{m}_g=\infty$. The resulting proton mass is 

\begin{equation}
m_p  = \overline{m_u+m_u+m_d+m_g}=
16523 {\hspace{0.2cm} {\rm (preon \hspace{0.2cm} mass \hspace{0.2cm} units)}.}
\label{eq:pmass}
\end{equation}

\noindent
We can convert $m_\mu$ and other particle masses from preon mass 
units into proton mass units, $m_p$ by dividing these masses by the quantity
(\ref{eq:pmass}).
The computed masses of the composite fermions 
[converted into the $m_p$-units  with the use of (\ref{eq:pmass})] 
are presented in Table \ref{t:massresult}, where the experimental
fermion masses \cite{properties} are listed in the last column 
for comparison (expressed also in units of the proton mass).
\begin{table}[htb]
\caption{Predicted and experimental rest masses of quarks and leptons}
\label{t:massresult}
\begin{center}
\small
\begin{tabular}
{|cc|c|c|c|c|} \hline
\multicolumn{2}{|c|}{Particle and} & {Number of } & Computed & Mass & Experimental \\
\multicolumn{2}{|c|}{its structure} & {preons in each} & total mass &  converted & mass \cite{properties} \\
\multicolumn{2}{|c|}{} &  { component } & (preon units) & into $m_p$ & ($m_p$) \\ \hline
\multicolumn{6}{|c|} {First generation} \\ 
$e^-$ & $3y$ & 9 & 9 & 0.0005447 & 0.0005446 \\ 
$\nu_e$ & $6y\overline{y}$ & 36 & 0 & 0 & - \\
$u$ & $ y_1 +  y_1 $ & 78  & 78 & 0.00472 & 0.001 to 0.005 \\
$d$ & \ \ $u + \nu_e e^-$ & 123 & 123 & 0.007443 & 0.003 to 0.009 \\
\hline
\multicolumn{6}{|c|}{Second generation} \\
$\mu^-$ & $\nu_e e^- \nu_\mu$ & $\overline{48+39}$ & 1872.6778 & 0.1133 & 0.1126 \\
$\nu_\mu $ & $y_1 \nu_e \overline{y}_1$ & 114  & 0 & 0 & -  \\
$c$ & $ y_2 + y_2 $ & $\overline{165+165}$ & 27225 & 1.6477 & 1.57 to 1.9  \\
$s$ & \ \ $c+e^-$ & $\overline{165+165+9}$ & 2751 & 0.1665 & 0.11 to 0.19 \\
\hline
\multicolumn{6}{|c|}{Third generation} \\
$\tau^-$ & $\nu_\mu \mu^- \nu_\tau $ & $\overline{156+201}$ & 31356 & 1.8977
 & 1.8939 \\
$\nu_\tau$ & $u \nu_e \overline{u}$ & 192  & 0 & 0 & - \\
$t$ & $ y_3 +  y_3$ & $\overline{1767+1767}$ & 3122289 & 188.96 
& $189.7\pm 5$ \\
$b$ & \ \ $t+\mu^-$ & $\overline{1767+1767+48+39}$ & 76061.5 & 4.60 & 4.2 to 4.7  \\
\hline
\end{tabular}
\end{center}
\end{table}
This Table illustrates the family-to-family similarities in the particle
structures. For instance, in each family, the $d$-like quark is a combination of the
$u$-like quark, with a charged lepton belonging to the lighter family.
Charged leptons appear as combinations of the neutrino from the
same family with the neutrinos and charged leptons from the lighter
family. 
It is conceivable that ring-closed strings,
similar to that of the the electron neutrino,
may appear on higher structural levels 
of the hierarchy, which could be regarded as ``heavy neutrinos'',
$\nu_h =6y_1\overline{y}_1$. They
can further form ``ultra-heavy'' neutrinos
$\nu_{uh} =3(\overline{y}_1 \nu_h u)e^-$, and so on.
With these neutral structures 
the components $y_2$ and $y_3$ of the $c$- and $t$-quarks 
can be written as: 
$y_2=u\nu_e u\nu_e e^-$ and $y_3= \nu_{uh}y$.

The experimental masses of the charged leptons are known with 
high accuracy. They are given in the last column of 
Table \ref{t:massresult} for comparison, so that 
we can estimate the accuracy of our model to be 
of about 0.5\% (only the experimental masses of the charged
leptons are used for the comparison). A higher accuracy could be achieved by 
taking into account some extra factors contributing to the 
masses, such as binding energies and/or the (non-vanishing) 
neutrino masses.

\section{Conclusions}
Our model explains the origin of the observed variety of elementary
particles and reproduces the masses of three generations of the fundamental 
fermions without using input parameters. The accuracy of this model
is estimated to be of about 0.5\%.

\end{document}